\documentclass[journal]{IEEEtran}
\ifCLASSINFOpdf
\else
\fi
\usepackage{amssymb}
\usepackage{amsmath}
\usepackage{bm}
\usepackage{mathrsfs}
\usepackage{algorithm}
\usepackage{algpseudocode}
\usepackage{graphicx}
\usepackage{subfigure}
\usepackage{tabularx}
\usepackage{dsfont}
\usepackage{multirow}
\newtheorem{remark}{Remark}
\newtheorem{define}{Definition}
\usepackage{booktabs}
\usepackage[graphicx]{realboxes}
\begin{document}
%
\title{Reconstruction of Delay Differential Equation via Learning Parameterized Dictionary}
%
%
%

\author{Yuqiang~Wu
\thanks{Y. Wu is with School of Artificial Intelligence and Automation, Huazhong University of Science and Techonology,
Wuhan, Hubei, 430074 China.
e-mail: (wuyuqiang@hust.edu.cn).}}
\maketitle

\begin{abstract}
This paper presents a variant of sparse representation modeling method,
which has a promising performance of reconstruction of delay differential equation from sampling data.
In the new method, a parameterized dictionary of candidate functions is constructed
against the traditional expanded dictionary.
The parameterized dictionary uses a function with variables to represent a series of functions.
It accordingly has the ability to express functions in the continuous function space
so that the dimension of the dictionary can be exponentially decreased.
This is particularly important when
an exhaustion of candidate functions is needed to construct appropriate dictionary.
The reconstruction of delay differential equation is such the case
that each possible delay item should be considered as the basis to construct the dictionary
and this naturally induces the curse of dimensionality.
Correspondingly, the parameterized dictionary uses a variable to model the delay item
so the curse disappears.
Based on the parameterized dictionary, the reconstruction problem is then rewritten and treated as a
mixed-integer nonlinear programming with both binary and continuous variables.
To the best of our knowledge, such optimization problem is hard to solve with the traditional mathematical methods
while the emerging evolutionary computation provides competitive solutions.
Hence, the evolutionary computation technique is considered
and a typical algorithm named particle swarm optimization is adopted in this paper.
Experiments are carried out in 5 test systems
including 3 well-known chaotic delay differential equations such as Mackey-Glass system.
The experiment result shows the effectiveness of the new method to
reconstruct delay differential equation.
\end{abstract}

\begin{IEEEkeywords}
Reconstruction, delay differential equation,
parameterized dictionary, mixed-integer nonlinear programming, evolutionary computation,
particle swarm optimization.
\end{IEEEkeywords}

%
\IEEEpeerreviewmaketitle

\section{Introduction}
%
%
%
%
\IEEEPARstart{R}{econstruction} of system dynamics from measurement data is a longstanding interest topic
in physics\cite{wang2016data}.
Reconstruction problem, also known as the inverse engineering, exists in a wide range of physical systems.
Time delay systems\cite{voss1997reconstruction,PhysRevLett.81.558,PhysRevE.64.056216,wang2012reverse},
stochastic systems\cite{PhysRevE.72.026202},
partial differential equations systems\cite{rudy2017data,li2019sparse} and
networks\cite{PhysRevLett.114.028701}are some of the most significant and challenging ones.
Various types of information are distilled from the unknown system, such as fractal dimensions, Lyapunov exponents, entropy and governing equations\cite{bradley2015nonlinear,brunton2016discovering},
aiming at describing and understanding the system.
Many data-driven reconstruction methods have been applied in this field such as embedding\cite{kantz2004nonlinear},
symbolic regression\cite{schmidt2009distilling}, statistical inference\cite{PhysRevE.97.022301}, etc.
Among all these methods, sparse representation modelling is considered the most promising one because it brings simplicity and interpretability\cite{wang2011predicting,wang2011network,shen2014reconstructing}.
In the sparse representation modelling, dictionary of possible system dynamic items is constructed with prior knowledge
and regularization is adopted to introduce sparsity.
To establish such model, Least absolute shrinkage and selection operator(LASSO\cite{tibshirani2015statistical}),
sparse Bayesian learning(SBL)\cite{pan2016sparse} and multi-objective evolutionary algorithms(MOEA\cite{li2014evolutionary}) are applied and proved effective.\par
Although sparse representation modeling is now considered the paradigm of the reconstruction problem,
it has limited capacity in the reconstruction of delay differential equation(DDE) due to the two inevitable
challenges as illustrated in the following:\par
\begin{itemize}
\item \textbf{The curse of dimensionality of the dictionary.}
The candidate dictionary must be designed as complete and accurate enough as possible to contain the true dynamics items,
specifically the delay items in DDE, according to our prior knowledge of the system.
For example, the dictionary of a simple system with the formulation of
$\dot{x}(t)=x(t-1.5)+x^2$ is expected to be $\bm{\Theta}(x)=\begin{bmatrix} 1,x,x(t-0.1),\ldots,x(t-1.5),x^2,x^2(t-0.1),\ldots)\end{bmatrix}$, or more complicated.
Obviously, enhancing the degree of discrete delay exponentially increases the dimension of the dictionary.
As is known, high dimensional dictionary matrix is expensive in both hardware storage and software computation,
thus the curse of dimensionality occurs.
Therefore, it is unrealistic to construct an appropriate dictionary under the traditional framework.
\end{itemize}
\begin{itemize}
\item \textbf{Unknown of the sparsity level.}
A proper value of the sparsity controller,
hyper-parameter $\lambda$ could only be determined by brute-force search.
The exact reconstruction fails when $\lambda$ is inappropriate.
To make matters worse, there may exists no right $\lambda$ if the right system items are not included in the dictionary.
\end{itemize}
\par
Above obstacles make the reconstruction of DDE an unfinished question.
Aiming at an exact reconstruction of DDE, a variant of sparse representation modeling method is
proposed in this paper.
The contributions of the paper is illustrated as follows.
Firstly, a parameterized dictionary expressing candidate functions in a low-dimensional
function space is novelly presented, for getting rid of the curse of dimensionality.
Correspondingly, the formulation of the reconstruction problem is rewritten
as the mixed-integer nonlinear programming(MINLP).
Secondly, the emerging optimization technique named evolutionary computation(EC)
is introduced to the reconstruction problem.
Thirdly, the effectiveness of the proposed method is validated in 5 test systems
containing 3 well-known chaotic DDE.\par
The remainder of this paper is organized as follows. Section II introduces the proposed method and the reformulation of the reconstruction problem.
Section III illustrates the EC and the details of the algorithm.
Section IV presents the experiments and relating results.
Section V provides an important technical discussion of the new method.
Section VI concludes this paper.\par
\section{Problem Formulation}
The proposed method is a variant of sparse representation modelling. Hence, the sparse representation modelling is briefly reviewed in the beginning.
\subsection{Sparse representation modelling}
Consider a system governed by delay differential equations(DDEs) as:
\begin{equation}
\frac{d}{dt}\bm{x}(t)=\bm{f}[\bm{x}(t),\bm{x}(t-\bm{\tau})],
\end{equation}
where $\bm{x}\in\mathds{R}^n$ represents the system state, $\bm{f}$ is the unknown system dynamics and $\bm{\tau}$ stands for the time-delays.
First, the measurement data $\bm{x}(t)$ is collected at the sampling time $t_1$ to $t_m$
and the derivative $\bm{\dot{x}}(t)$ is approximated through numerical difference.
Then, the dictionary $\bm{\Theta}(\bm{x})$ is constructed which contains possible items of $\bm{f}$ according to the prior knowledge of the system.
For example, a dictionary may consist of constant, polynomial and time-delay items:
\begin{equation}
\bm{\Theta}(\bm{x})=\begin{bmatrix}
                \bm{1} & \bm{x} & \bm{x}^2 & \bm{...} & \bm{x}(t-\tau_1) & \bm{x}(t-\tau_2) & \bm{...}\\
            \end{bmatrix}.
\end{equation}
After that, the sparse coefficients matrix $\bm{\Xi}=\begin{bmatrix}\bm{\xi_1}\;\bm{\xi_2}\;\bm{...}\;\bm{\xi_n}\end{bmatrix}$ is defined in which $\bm{\xi_i}$ is a sparse vector.
Thus the sparse regression problem is established as:
\begin{equation}
\bm{\dot{X}}=\bm{\Theta}(\bm{X})\bm{\Xi},
\end{equation}
where $\bm{\dot{X}}$ and $\bm{X}$ are $m\times n$ matrix as:
\begin{equation}
\bm{\dot{X}}\!=\!\begin{bmatrix}
                \dot{x}_1(t_1) & \cdots & \dot{x}_n(t_1)\\
                \vdots    & \ddots & \vdots  \\
                \dot{x}_1(t_m) & \cdots & \dot{x}_n(t_m)\\
            \end{bmatrix},
\bm{X}\!=\!\begin{bmatrix}
                x_1(t_1) & \cdots & x_n(t_1)\\
                \vdots     & \ddots & \vdots  \\
                x_1(t_m)& \cdots & x_n(t_m)\\
            \end{bmatrix}.
\end{equation}
This problem can be divided into $n$ independent optimization subproblems as:
\begin{equation}
\bm{\xi_{i}}^*=\arg\min\|\bm{\Theta(X)\xi_{i}}-\bm{\dot{X_{i}}}\|_{2}+\lambda\|\bm{\xi_{i}}\|_{0}\quad i=1,2,...,n,
\end{equation}
where $\bm{\xi_{i}}$ and $\bm{X_{i}}$ are the $i$th column of $\bm{\Xi}$ and $\bm{X}$.
$\lambda$ is the regularization hyper-parameter and the subscript 2 and 0 stands for $L_2$ and $L_0$ norm.
The solution of Eq.(5) is the sparse coefficients matrix $\bm{\Xi}^*$, thus we obtain the sparse
representation of the system and finish the reconstruction.\par
\subsection{Proposed method}
As illustrated in the introduction, sparse representation modelling has limitation in the reconstruction of DDE.
A new method is proposed based on sparse representation modelling,
which has a different formulation of the reconstruction problem.\par
\subsubsection{Parameterized dictionary}
A parameterized dictionary is novelly presented. Its definition and relating analysis are introduced in this section.
\begin{define}
$\bm{p_i}=[p_{i1},p_{i2},\ldots,p_{in_{p}}]^T$. $n_{p}$ is the dimension of $\bm{p_i}$.
\end{define}
\begin{define}
For $\bm{x}\in\mathds{R}^n$, $g(\bm{p_i})=x_{1}^{p_{i1}}(t-p_{i2})x_{2}^{p_{i3}}(t-p_{i4})\cdots x_{n}^{p_{i2n-1}}(t-p_{i2n})\cdots$.
\end{define}
\begin{define}
$\bm{p}=[\bm{p_1},\bm{p_2},\ldots,\bm{p_M}^T]$. $M$ is a given number which represents the max number of the reconstruction items.
$\bm{p}$ is the parameters vector of the dictionary which satisfies $\bm{p}\in\mathds{R}^{Mn_{p}}$.
\end{define}
\par The parameterized dictionary is constructed as:
\begin{equation}
\bm{\Theta}(\bm{x,p})=\begin{bmatrix}
                g(\bm{p_1}),g(\bm{p_2}),\ldots,g(\bm{p_M})
            \end{bmatrix}^T.
\end{equation}
It is apparent that the key to construct a parameterized dictionary is to construct $g(\bm{p_i})$ and determine $M$.\par
There are two advantages of the parameterized dictionary.
\begin{itemize}
\item It avoids the expansion of the detailed candidate functions through compressing them into the parameters.
As a consequence, the function space of the dictionary can be expressed roughly large
but still keep low-dimensional property.
In the meantime, the accuracy problem of the dictionary disappears because the parameters can be continuous.
\end{itemize}
\begin{itemize}
\item Simplicity and interpretability of the reconstruction system are obtained without the introduction of sparsity because $M$ is an artificially set number according to the prior knowledge and can be adjusted as user's wish.
Since $M$ is the max number of the reconstruction items, it doesn't need tuning once it is given.
\end{itemize}
\begin{remark}
For better explanation of the parameterized dictionary, a system with the formulation of $\dot{x}(t)=x(t-1.5)+x^2$ is analysed as the example.
Define $g(\bm{p_i})=x^{p_{i1}}(t-p_{i2})$ and set $M$. Its parameterized dictionary is expressed as
$\bm{\Theta}(x,\bm{p})=[g(\bm{p_1}),g(\bm{p_2}),\ldots,g(\bm{p_M})]^T$,
or expanded as
$\bm{\Theta}(x,\bm{p})=\begin{bmatrix}
                x^{p_{11}}(t-p_{12}),\ldots,x^{p_{M1}}(t-p_{M2})
            \end{bmatrix}^T.$
\end{remark}
\subsubsection{MINLP formulation}
Based on the parameterized dictionary, reconstruction problem can be formulated as:
\begin{equation}
\{\bm{\xi_{i}}^*,\bm{P_i}^*\}=\arg\min\|\bm{\Theta}(\bm{X,P_i})\bm{\xi_{i}}-\bm{\dot{X_{i}}}\|_{2}\quad\, i=1,2,...,n,
\end{equation}
where $\bm{P_i}$ is the $i$th column of the parameter variables matrix $\bm{P}$ and $\bm{P}\in\mathds{R}^{Mn_p\times n}$.
Note that regularization is not used in Eq.(7), so the hyper-parameter tuning problem no more exists.\par
Eq.(7) is a non-convex optimization problem and contains continuous and integer variables. A feasible idea is to reformulate it as mixed-integer nonlinear programming.
Without loss of generality, consider an $n$-dimensional system, in which
$\bm{\xi_i}$ is expanded as $\begin{bmatrix} \xi_{i1}\;\xi_{i2}\;0\;...\;\xi_{il}\;0\;...\;\xi_{iM}\end{bmatrix}^T$.
\begin{define}
The simplified coefficients vector $\bm{d_i}$ is defined as $\bm{d_i}=\begin{bmatrix} 1,1,0,...,1,0,...,1\end{bmatrix}^T$ which satisfies $\bm{\xi_i}=\begin{bmatrix} \xi_{i1},\xi_{i2},0,...,\xi_{il},0,...,\xi_{iM}\end{bmatrix}^T\circ\bm{d_i}$, where the operator "$\circ$"
is the Hadamard product of the matrix.
0 and 1 in $\bm{d_i}$ stands for the zero and non-zero items in $\bm{\xi_i}$.
The simplified coefficients matrix $\bm{D}$ is defined as the combination of all $\bm{d_i}$.
\end{define}\par
Hence, the reconstruction problem is transformed into mixed-integer nonlinear programming which is formulated as:
\begin{equation}
\{\!\bm{d_i}^*,\bm{P_i}^*,\bm{\xi_i}^*\}=\arg\min\|\bm{\Theta}(\bm{X,P_i})(\bm{d_i}\circ\bm{\xi_i})-\bm{\dot{X}_i}\|_2\ i=1,2,...,n.
\end{equation}\par
A solution of Eq.(8) is generated with three steps. Firstly, determine $\bm{d_i}$
which represents the trade-off of the dictionary items. Secondly, determine $\bm{P_i}$.
Thirdly, perform least square method to obtain $\bm{\xi_i}$.
Thus, the MINLP problem can be treated in a bi-level optimization framework. In detail,
the outside optimization aims to find the optimal $\bm{d_i}$ while
the inside optimization searches for the optimal $\bm{P_i}$ and its relating $\bm{\xi_i}$.
It is clear that the outside optimizes the binary variables and the inside optimizes the continuous variables.\par
\section{Proposed Algorithm}
Above bi-level optimization problem is an NP-hard problem with high nonlinearity.
It cannot be efficiently solved by traditional mathematical methods.
However, an emerging optimization method named evolutionary computation has the potential to
obtain solutions with high quality and acceptable computation cost.
Therefore, EC is employed in both outside and inside optimization and it is introduced in the beginning as preliminary of the proposed algorithm.\par
\subsection{Evolutionary computation}
Evolutionary computation\cite{salcedo2016modern,del2019bio,fausto2020ants} represents a class of nature-inspired optimization algorithms.
It aims at global optimization and works in the absence of explicit problem formulation and gradient information. As a consequence,
it has a broad application in many scientific and engineering problems\cite{gotmare2017swarm,darwish2019survey} where traditional mathematical methods fail.
To emphasis, combinational optimization\cite{liefooghe2018evolutionary,ramos2020metaheuristics} and multi-modal optimization\cite{das2011real,li2016seeking} are some of the most important applications in EC field.
A lot of evolutionary algorithms(EAs) have been presented and well studied. Among various algorithms,
particle swarm optimization(PSO)\cite{shi1998modified,bonyadi2017particle}
gains special attention as it has strong global optimization ability and is easy to realize.
Therefore, PSO is adopted in this paper and introduced here to explain the mechanism of EA.\par
\begin{figure}[ht]
\centering
\includegraphics[width=7cm,height=12cm]{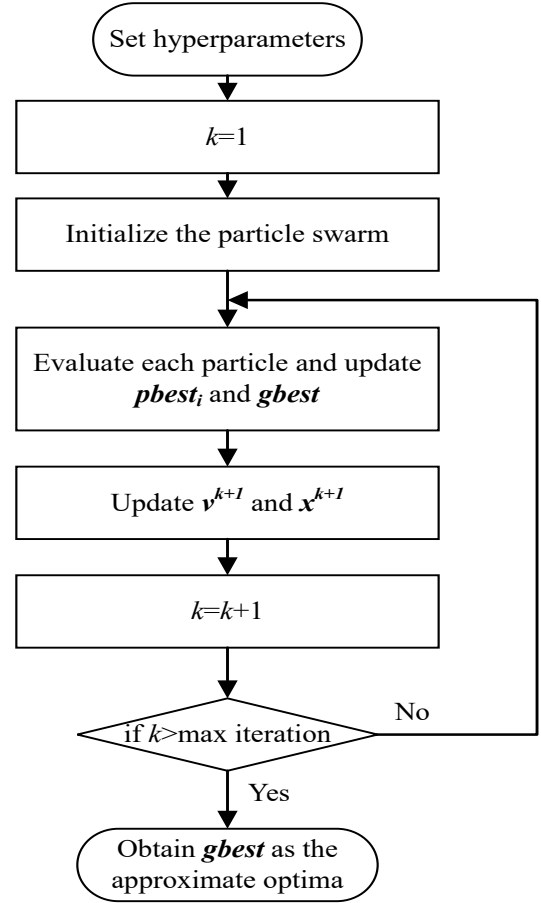}
\caption{Flowchart of PSO.}
\end{figure}
In PSO, particles (or individuals in other EAs) are the basic units of optimization.
Each particle has two characteristics, which are position $\bm{x_i}$ and velocity $\bm{v_i}$.
$\bm{x_i}$ represents the solution in optimization and $\bm{v_i}$ represents the search direction and step size.
Firstly, $N$ particles are randomly initialized with $\bm{x_i^k}$ and velocity $\bm{v_i^k}$,
in which $k$ means the iteration number.
Then, the objective value of each particle is evaluated.
$\bm{pbest_i}$ and $\bm{gbest}$ are defined which stand for the best position a particle finds in its own search history
and the best position the particle swarm finds in the whole search history.
Thus, $\bm{pbest_i}$ and $\bm{gbest}$ in the present iteration can be obtained after evaluation.
Next, particles are updated with velocities and positions according to the rule as:
\begin{equation}
\left\{
\begin{array}{lr}
\bm{v_i^{k+1}}=\omega\bm{v_i^k}+c_1\bm{r_1}\circ(\bm{pbest_i}-\bm{x_i^k})+c_2\bm{r_2}\circ(\bm{gbest}-\bm{x_i^k})\\
 \bm{x_i^{k+1}}=\bm{x_i^k}+\bm{v_i^{k+1}}
 \end{array},
\right.
\end{equation}
where $\omega$, $c_1$ and $c_2$ are the hyper-parameters which are often defaults and
$\bm{r_1}$, $\bm{r_2}$ are random vectors uniformly distributed in [0,1].
The operator "$\circ$" is the Hadamard product of the matrix.
At this point, an iteration is over and the loop continues until the iteration comes to the max iteration as set.
$\bm{gbest}$ in the last iteration means the best solution found of the optimization problem.
Note that $\bm{gbest}$ is not equivalent to the global optimum although it is always a competitive solution.
The flowchart of PSO is shown in Fig.1. An intuitive description of the optimization process is shown in Fig.2.\par
\begin{figure}[ht]
\centering
\subfigure[]{\includegraphics[width=4cm,height=4cm]{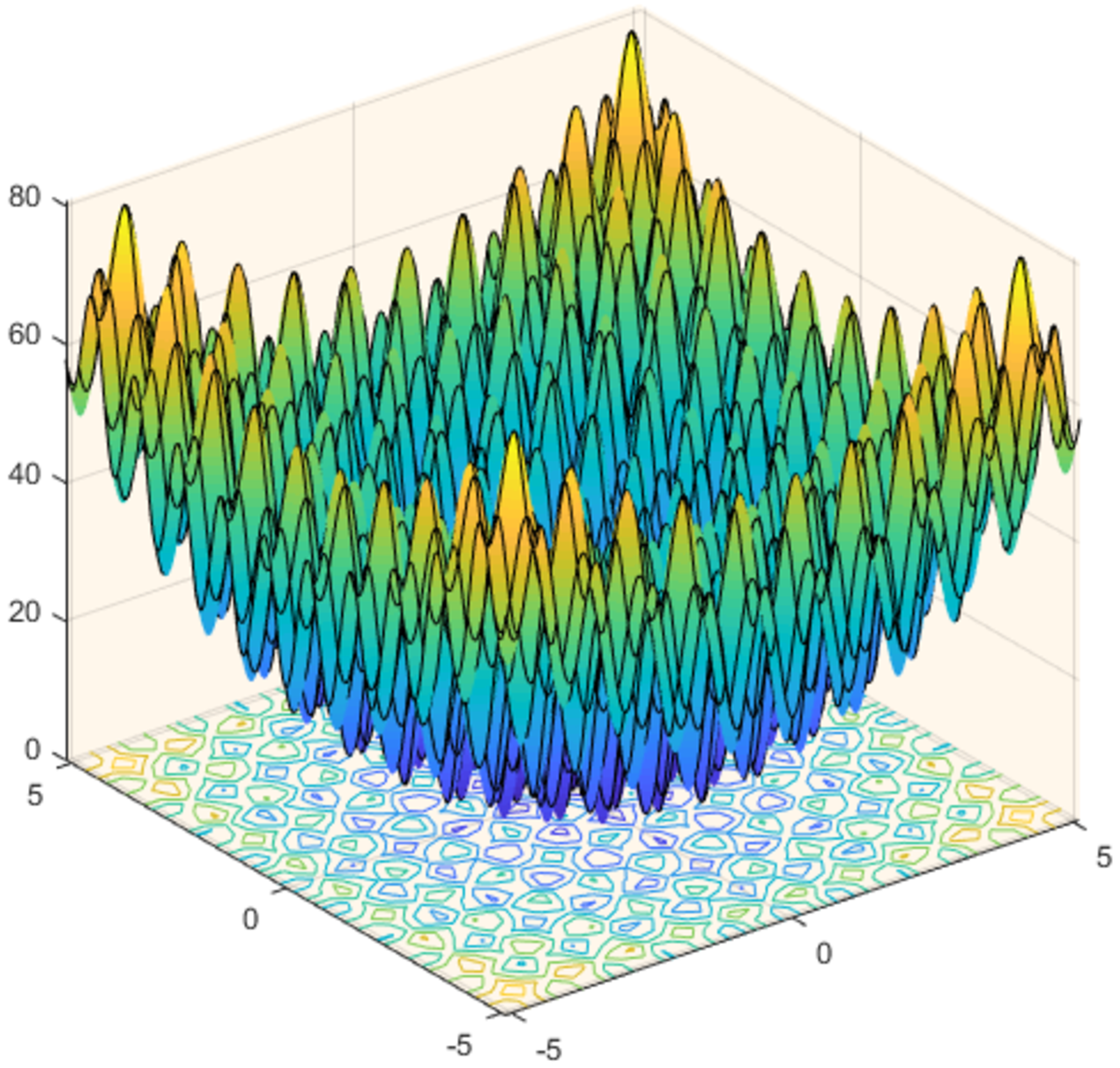}}
\subfigure[]{\includegraphics[width=4cm,height=4cm]{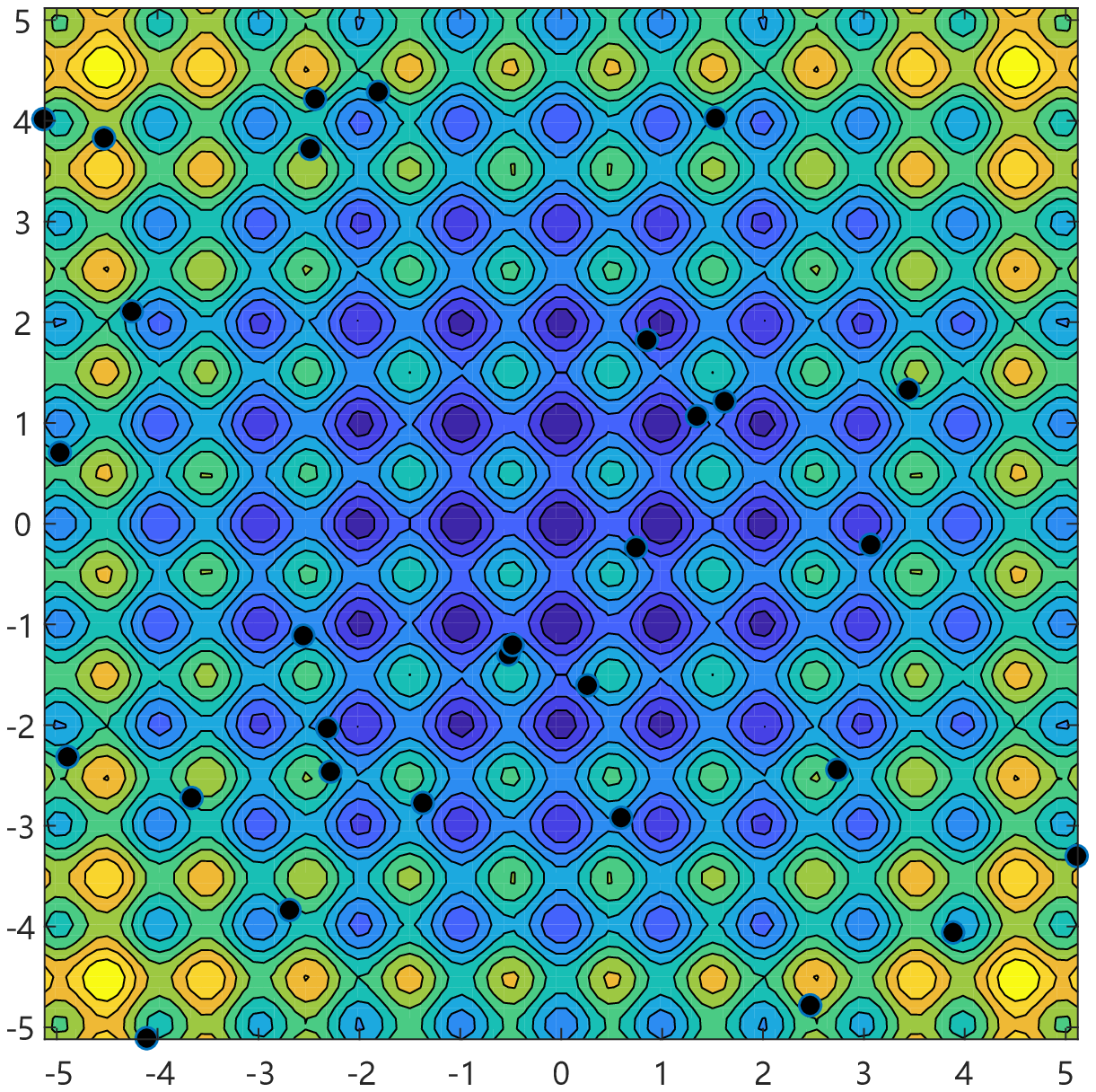}}
\subfigure[]{\includegraphics[width=4cm,height=4cm]{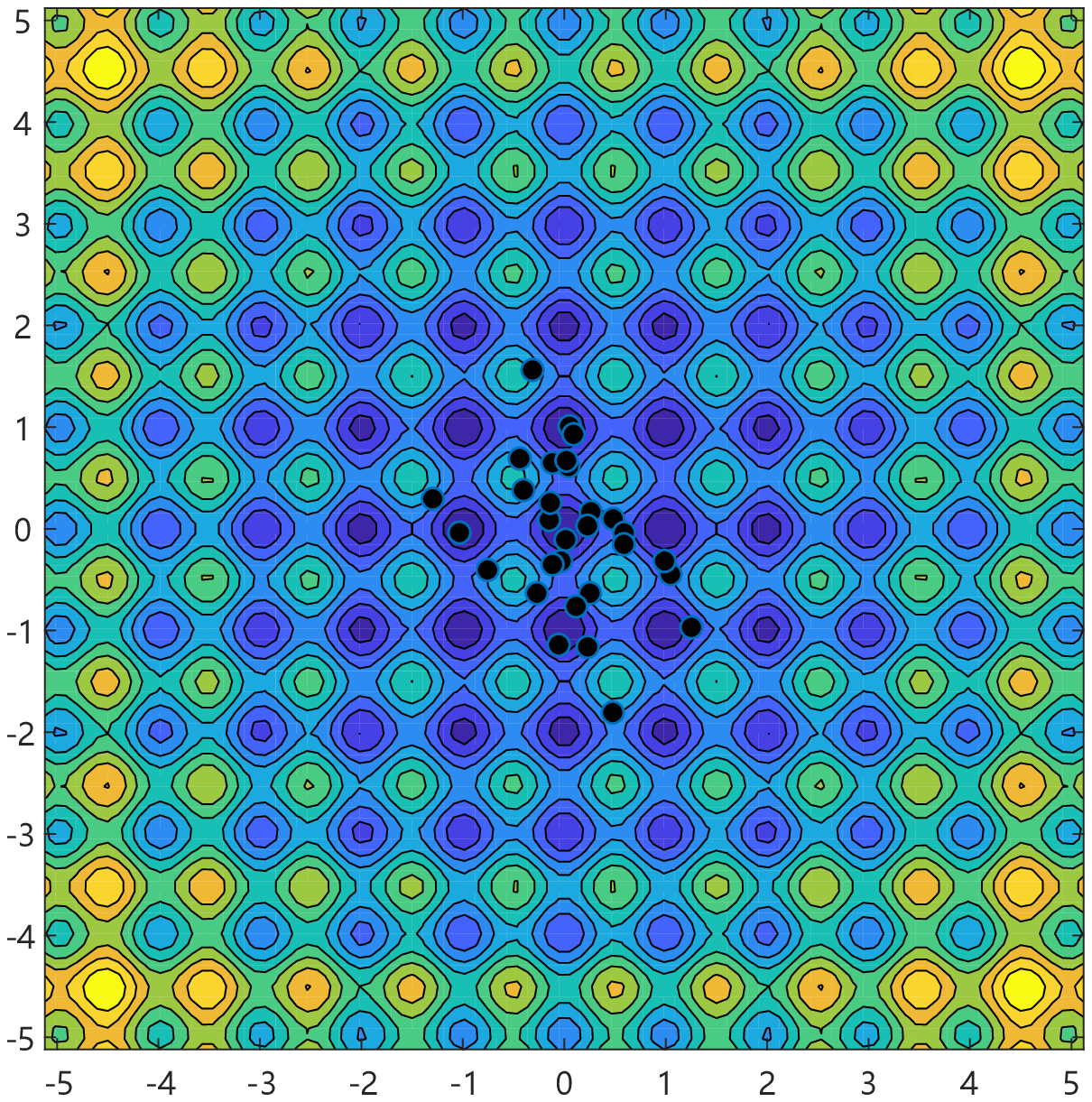}}
\subfigure[]{\includegraphics[width=4cm,height=4cm]{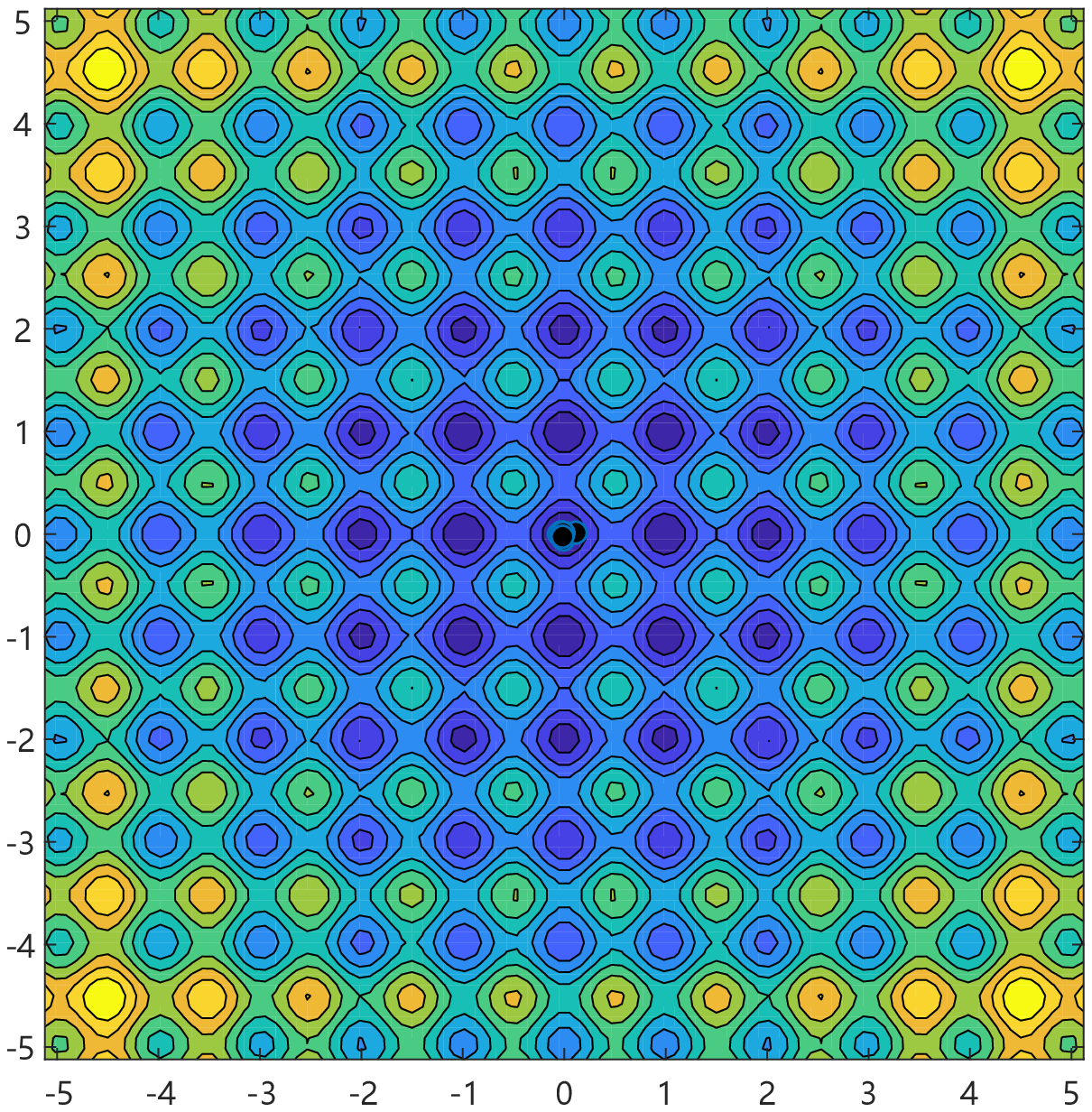}}
\caption{(a) is the visualization of a 2-dimensional Rastrigin¡¯s function with the formulation of $f(\bm{x})=\sum_{i=1}^2(x_i^2-10cos(2\pi x_i)+20)$ and
the global optimum is $[0,0]^T$. (b)-(d) are the visualization of the optimization process. The function is plotted in contours and the black dots are the particles.
(b) is the $k=1$ snap shot(randomly initialization). (c) is the $k=25$ snap shot. (d) is the $k=100$ snap shot. It shows the particles converge to the global optimum after a certain number of iterations.}
\end{figure}
Although PSO is firstly proposed and commonly used in the continuous optimization problem. It is worth noting that there are binary versions of PSO\cite{kennedy1997discrete,khanesar2007novel,mirjalili2013s}
for the optimization problem defined in binary space.
Through adding a transfer function mapping the continuous and binary solution space\cite{mirjalili2013s},
the continuous version of PSO can be conveniently transferred into the binary one.
Due to this consistency of PSO mechanism in two kinds of versions,
it is considered unnecessary to introduce the binary version of PSO alone in this paper.
\subsection{Algorithm}
To solve the optimization problem of Eq.(8), the binary version and the continuous version of PSO are respectively employed in the outside and the inside optimization,
which are denoted as $PSO_{out}$ and $PSO_{in}$ in this paper.
Consequently, the whole algorithm to solve the reconstruction problem is described as follow:
\begin{algorithm}[H]
\caption{Proposed algorithm}
\begin{algorithmic}[1]
\State {Construct $\bm{\Theta}(\bm{x,p}$) and set $M$;}
\State {Set the hyper-parameters of $PSO_{out}$ and $PSO_{in}$ including $N_{out}$ and $N_{in}$, $I_{max}^{out}$ and $I_{max}^{in}$, etc.;}
\For {each $i=1,2,...,n$}
\State {Initialize the population $\bm{d_i^{N_{out}}}$;}
\While{iteration $\leq$ $I_{max}^{out}$}
\For {each particle of $\bm{d_i^{N_{out}}}$}
\State {Initialize the population $\bm{P_i^{N_{in}}}$;}
\While{iteration $\leq$ $I_{max}^{in}$}
\For {each particle of $\bm{P_i^{N_{in}}}$}
\State {Evaluate Eq.(8);}
\State {Update $\bm{P_i}$;}
\EndFor
\EndWhile
\State {Obtain $\bm{P_i^*}$ and $\bm{\xi_i^*}$ of $\bm{P_i^{N_{in}}}$ for $\bm{d_i}$;}
\State {Update $\bm{d_i}$;}
\EndFor
\EndWhile
\State {Obtain $\bm{d_i^*}$, $\bm{P_i^*}$ and $\bm{\xi_i^*}$;}
\EndFor
\State{Combine $\bm{d_i^*}$, $\bm{P_i^*}$ and $\bm{\xi_i^*}$ thus obtain $\bm{D^*}$, $\bm{P^*}$ and $\bm{\Xi^*}$. }
\end{algorithmic}
\end{algorithm}
\section{Experiments and Results}
\subsection{Experiment settings}
\begin{table}[]
\newcommand{\tabincell}[2]{\begin{tabular}{@{}#1@{}}#2\end{tabular}}
 \footnotesize
 \centering
 \caption{Characteristics description of 5 test systems.}
 \Rotatebox{90}{
\begin{tabular}{p{5pt}<{\centering}p{100pt}<{\centering}p{160pt}<{\centering}p{150pt}<{\centering}p{100pt}<{\centering}}
\toprule
\textbf{ID} & \textbf{Name} & \textbf{Formulation} & \textbf{Dictionary} & \textbf{Solution space} \\
\midrule
 \textbf{1} &\textbf{Linear system} & $\frac{d}{dt}\begin{bmatrix}  x\\ y\\z\\ \end{bmatrix}=\begin{bmatrix}  -0.1&2&0\\ -2&-0.1&0\\0&0&-0.3\\ \end{bmatrix}\begin{bmatrix}  x\\ y\\z\\ \end{bmatrix}$ &$g(\bm{p_i})=x^{p_{i1}}y^{p_{i2}}z^{p_{i3}},M=5$ & $p_{ij}\in[0, 5],p_{ij}\in\mathds{Z}$ \\ \hline
 \textbf{2} &\textbf{Lorenz} & $\left\{\begin{array}{lr}\dot{x}=\sigma(y-x)\\\dot{y}=x(\rho-z)-y\\\dot{z}=xy-\beta z\end{array}\right.,
 \begin{matrix}\sigma=10\\ \rho=28\\ \beta=8/3\end{matrix}$ &  $g(\bm{p_i})=x^{p_{i1}}y^{p_{i2}}z^{p_{i3}},M=5$ & $p_{ij}\in[0, 5],p_{ij}\in\mathds{Z}$ \\ \hline
 \textbf{3} &\textbf{Delayed R\"{o}ssler}\cite{ghosh2008multiple} & \tabincell{c}{$\left\{\begin{array}{lr} \dot{x}=-y-z+\alpha_1x(t-\tau_1)+\alpha_2x(t-\tau_2)\\
  \dot{y}=x+\beta_1y\\ \dot{z}=\beta_2+z(x-\gamma)\end{array}\right.$\\
 $\alpha_1=0.2, \alpha_2=0.5, \beta_1=\beta_2=0.2$\\ $\gamma=5.7, \tau_1=1, \tau_2=2$ }& $g(\bm{p_i})=\begin{matrix}x^{p_{i1}}(t-p_{i2})\cdot\\y^{p_{i3}}(t-p_{i4})\cdot\\z^{p_{i5}}(t-p_{i6})\end{matrix},M=5$ & $p_{ij}\in[0, 5],p_{ij}\in\mathds{Z}$ \\ \hline
 \textbf{4} & \textbf{Ikeda}\cite{ikeda1987high} & \tabincell{c}{$\dot{x}(t)=-x(t)+\alpha\sin(x(t-\tau))$ \\ $\alpha=6,\tau=1.59$} & $g(\bm{p_i})= \begin{matrix}x^{p_{{i1}}}(t-p_{i2})\cdot\\ \sin^{p_{i3}}(x(t-p_{i4}))\end{matrix},M=5$ &
 \tabincell{c}{$p_{ij}\in[0, 5],$\\ $p_{i1,i3}\in\mathds{Z},p_{i2,i4}\in\mathds{R}$} \\   \hline
 \textbf{5} &\textbf{Mackey-Glass}\cite{mackey1977oscillation} & \tabincell{c}{$\dot{x}(t)=-bx(t)+\frac{ax(t-\tau)}{1+x^c(t-\tau)}$ \\ $a=0.2, b=0.1, c=10, \tau=20$}  &
 $g(\bm{p_i})=\frac{x^{p_{i1}}x^{p_{i2}}(t-p_{i5})}{x^{p_{i3}}+x^{p_{i4}}(t-p_{i5})},M=5$ & \tabincell{c}{$p_{i1,i2,i3,i4}\in[0, 10],$ \\$p_{i5}\in[10, 30],p_{ij}\in\mathds{Z}$} \\
 \bottomrule
 \end{tabular}
 }
 \end{table}
To show the effectiveness, the proposed method is executed in reconstruction problems of 5 systems.
The characteristics description of the test systems are listed in Table I.
System 1 and 2 are relatively simple systems that governed by ordinary differential equations(ODEs).
They can be treated as the simple version of DDEs with no delay.
System 3-5 are governed by well-known chaotic DDEs. Specifically, system 3 has two delays of different values,
system 4 has a delay with higher accuracy and system 5 has a fraction expression.\par
To establish the reconstruction problem, data is collected in the simulation system with the sampling interval of 0.01s
and is intercepted with the length of 20s in system 1-4 and 80s in system 5.
After that, the approximation of $\bm{\dot{x}}(t)$ is calculated by the center difference.
The construction of parameterized dictionary with the solution space of each system is shown in Table I.
It is notable that the time series of a single time-delay item like $\bm{x}(t-\tau)$
is obtained from the entire sampling data collection $\bm{x}(t)$.
Consequently, the real number field in the solution space of system 4
is physically realized with the accuracy of 0.01s.\par
Binary PSO\cite{kennedy1997discrete} and couple-based PSO\cite{wu2019couple} are selected as $PSO_{out}$ and $PSO_{in}$ in the experiment, which are respectively denoted as BPSO and CPSO.
BPSO is the first and also the most widely used version of binary PSO
while CPSO is a modified algorithm of continuous PSO which is designed for multi-modal optimization problem.
Their hyper-parameter settings are listed in Table II, in which
$M$ and $n_p$ are obtained directly from Table I.
Notably, the hyper-parameters are set in a general mode without tuning according to \cite{kennedy1997discrete} and \cite{wu2019couple}.
Finally, due to the stochastic feature of PSO, each tested case is run 100 times independently
and all experiments are realized in Matlab code. \par
 \begin{table}[h]
 \centering
 \caption{Hyper-parameter settings of BPSO and CPSO.}
 \begin{tabular}{p{20pt}<{\centering}p{20pt}<{\centering}p{20pt}<{\centering}p{20pt}<{\centering}p{10pt}<{\centering}p{10pt}<{\centering}p{10pt}<{\centering}p{10pt}<{\centering}p{10pt}<{\centering}p{10pt}<{\centering}}
     \toprule
     \multirow{2}{*}{\textbf{BPSO}}& $\bm{N_{out}}$ & $\bm{I_{max}^{out}}$ &$\bm{V_{lim}^{out}}$& \multicolumn{2}{c}{$\bm{\omega}$} & \multicolumn{2}{c}{$\bm{c_{1}}$} & \multicolumn{2}{c}{$\bm{c_{2}}$} \\ \cline{2-10}
   & $M$ & $N_{out}^2$ & 4 &\multicolumn{2}{c}{0.6}  & \multicolumn{2}{c}{2}& \multicolumn{2}{c}{2} \\ \hline
  \multirow{2}{*}{\textbf{CPSO}} &$\bm{N_{in}}$ & $\bm{I_{max}^{in}}$ &$\bm{V_{lim}^{in}}$& $\bm{\omega_a}$ & $\bm{\omega_b}$ & $\bm{c_{1a}}$ & $\bm{c_{1b}}$& $\bm{c_{2a}}$ & $\bm{c_{2b}}$ \\ \cline{2-10}
   & $Mn_p$  & $N_{in}^2$ & 0.6 &0.2 &0.3 &0.9 &0.3 &1.5 &1.5 \\
 \bottomrule
 \end{tabular}
 \end{table}
\subsection{Analysis of the results}
As there is no comparative work in the governing equation reconstruction of DDE systems,
only limited aspects of the experiment result are concentrated in this paper.
The reconstructed system formulation and the success ratio of each tested case are exhibited in Table III.
The criterion of a success reconstruction is the exact reconstruction,
which means any little deviation of the governing equation such as a reconstructed time delay 1.58 in system 4 is considered a fail.\par
  \begin{table}[]
 \footnotesize
 \centering
 \caption{Reconstruction results of 5 systems.}
 \begin{tabular}{p{2pt}<{\centering}p{180pt}<{\centering}p{50pt}<{\centering}}
\toprule
\textbf{ID} & \textbf{Reconstructed system} &\textbf{Success ratio} \\ \hline
   \textbf{1} & $\frac{d}{dt}\begin{bmatrix}  x\\ y\\z\\ \end{bmatrix}=\begin{bmatrix}  -0.1000&1.9999&0\\ -1.9999&-0.1000&0\\0&0&-0.3000\\ \end{bmatrix}\begin{bmatrix}  x\\ y\\z\\ \end{bmatrix}$  & 100/100\\ \hline
 \textbf{2}  & $\left\{\begin{array}{lr}\dot{x}=10.0301y-10.0545x\\\dot{y}=28.1237x-1.0042xz-1.0537y\\\dot{z}=1.0006xy-2.6657z\end{array}\right. $ & 100/100 \\ \hline
 \textbf{3}  & $\left\{\begin{array}{lr}\dot{x}=-0.9998y-0.9996z+\\ \quad\quad0.1996x(t-1)+0.5000x(t-2)\\
 \dot{y}=1.0000x+0.2000y\\\dot{z}=0.1931+0.9995xz-5.6958z\end{array}\right. $ & 85/100 \\ \hline
 \textbf{4} & $\dot{x}(t)=-0.9999x(t)+5.9983\sin(x(t-1.59))$  & 55/100\\  \hline
 \textbf{5} & $\dot{x}(t)=-0.0999x(t)+0.1999\frac{x(t-20)}{1+x^{10}(t-20)}$  & 8/100\\
 \bottomrule
 \end{tabular}
 \end{table}
 \begin{table}[]
 \centering
 \footnotesize
 \caption{The optimal objective values in experiment 4 and 5 from 30 times running. The bold number means it is a successful case.}
\begin{tabular}{|c|c|c|c|c|c|}
  \hline
  \textbf{Count} & \textbf{System 4} & \textbf{System 5} & \textbf{Count} & \textbf{System 4} & \textbf{System 5} \\ \hline
  1 & 0.0840 & 0.0262 & 16 & 0.0829& 0.0237 \\
  2 & \textbf{0.2398} & 0.0318 & 17 & \textbf{0.2501} & 0.0201 \\
  3 & 0.0972 & 0.0274 & 18 & \textbf{0.2368} & 0.0278 \\
  4 & 0.1013 & \textbf{0.0044} & 19 & \textbf{0.2476} & 0.0288 \\
  5 & \textbf{0.2468} & 0.0220 & 20 & \textbf{0.2383} & 0.0323 \\
  6 & \textbf{0.2434} & 0.0257 & 21 & 0.1003 & 0.0294 \\
  7 & 0.0558 & 0.0384 & 22 & \textbf{0.2390} & 0.0272 \\
  8 & \textbf{0.2589} & \textbf{0.0046} & 23 & \textbf{0.2376}& \textbf{0.0047} \\
  9 & \textbf{0.2500} & 0.0254 & 24 & \textbf{0.2576} & 0.0193 \\
  10 & \textbf{0.2440} & 0.0190 & 25 & \textbf{0.2369} & 0.0249 \\
  11 & 0.0430 & 0.0110 & 26 & \textbf{0.2445} & 0.0287 \\
  12 & 0.1013 & 0.0309 & 27 & \textbf{0.2488} & 0.0218 \\
  13 & \textbf{0.2487} & 0.0276 & 28 & 0.1038 & 0.0292 \\
  14 & \textbf{0.2424} & 0.0219 & 29 & \textbf{0.2438} & 0.0183 \\
  15 & \textbf{0.2377} & 0.0199 & 30 & \textbf{0.2508} & 0.0314 \\
  \hline
\end{tabular}
\end{table}
From Table III, it is shown that all systems are able to be exactly reconstructed and
ODE systems have the success ratio of 1. This validates the effectiveness of the proposed method. 
It is also observed that
DDE systems have a lower success ratio.
Combining Table II and Table III, it is shown that
the success ratio decreases with the increase of the dictionary complexity and the solution space.
In theoretical aspect, the nonlinearity of the optimization problem Eq.(8) becomes higher as the dictionary and the solution space become more complex, which makes
the algorithm more likely to trap into local optima.\par
Table IV lists the optimal objective values in experiment 4 and 5 of 30 running times.
It is shown that both successful and unsuccessful results have a small fitting error.
With the simplicity induced by the parameterized dictionary, it is further indicated that
the reconstructed governing equations achieve both high fitting accuracy and good generalization ability.
This is an important advantage of the proposed method.
Besides, an abnormality appears in Table IV where the successful cases in system 4 have a bigger
optimal objective value than the unsuccessful ones.
It tells that the right reconstruction is the local optimum of Eq.(8).
In other words, the algorithm is capable of finding the global optimum but the global optimum is not related to the true system dynamics.
From the perspective of theoretical analysis, this phenomenon is caused by the deviation of the center difference.
If the approximation of $\bm{\dot{x}}(t)$ could be more accurate through adopting better approximation method or improving the accuracy of sampling, the abnormality could be effectively moderated.\par
In addition, a visualisation of experiment 1-4 with the comparison of the original and the reconstructed system is shown in Fig. 3. It is intuitively observed that the systems are successfully reconstructed.\par
\begin{figure}[ht]
\centering
\subfigure[]{\includegraphics[width=4cm,height=4cm]{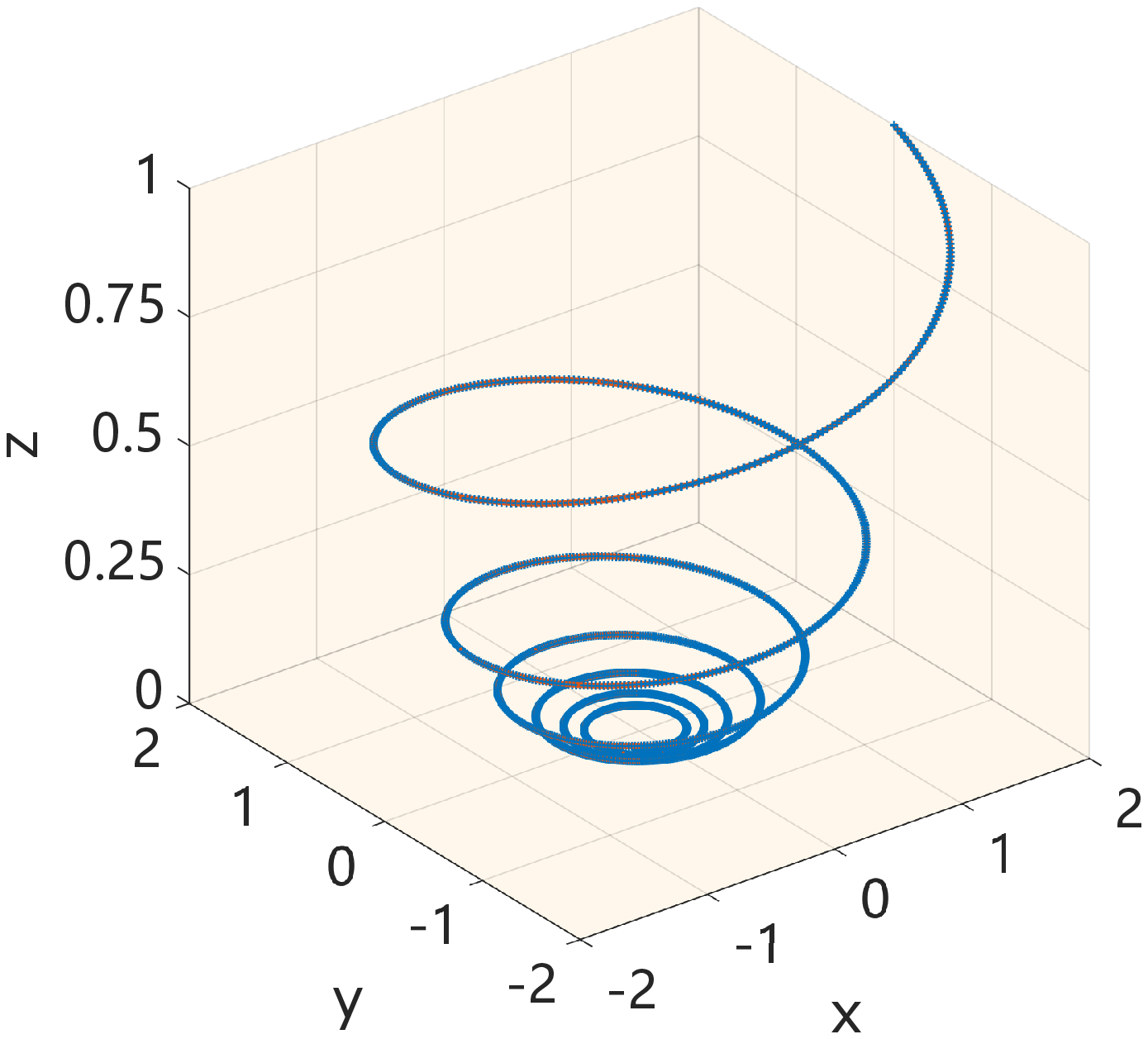}}
\subfigure[]{\includegraphics[width=4cm,height=4cm]{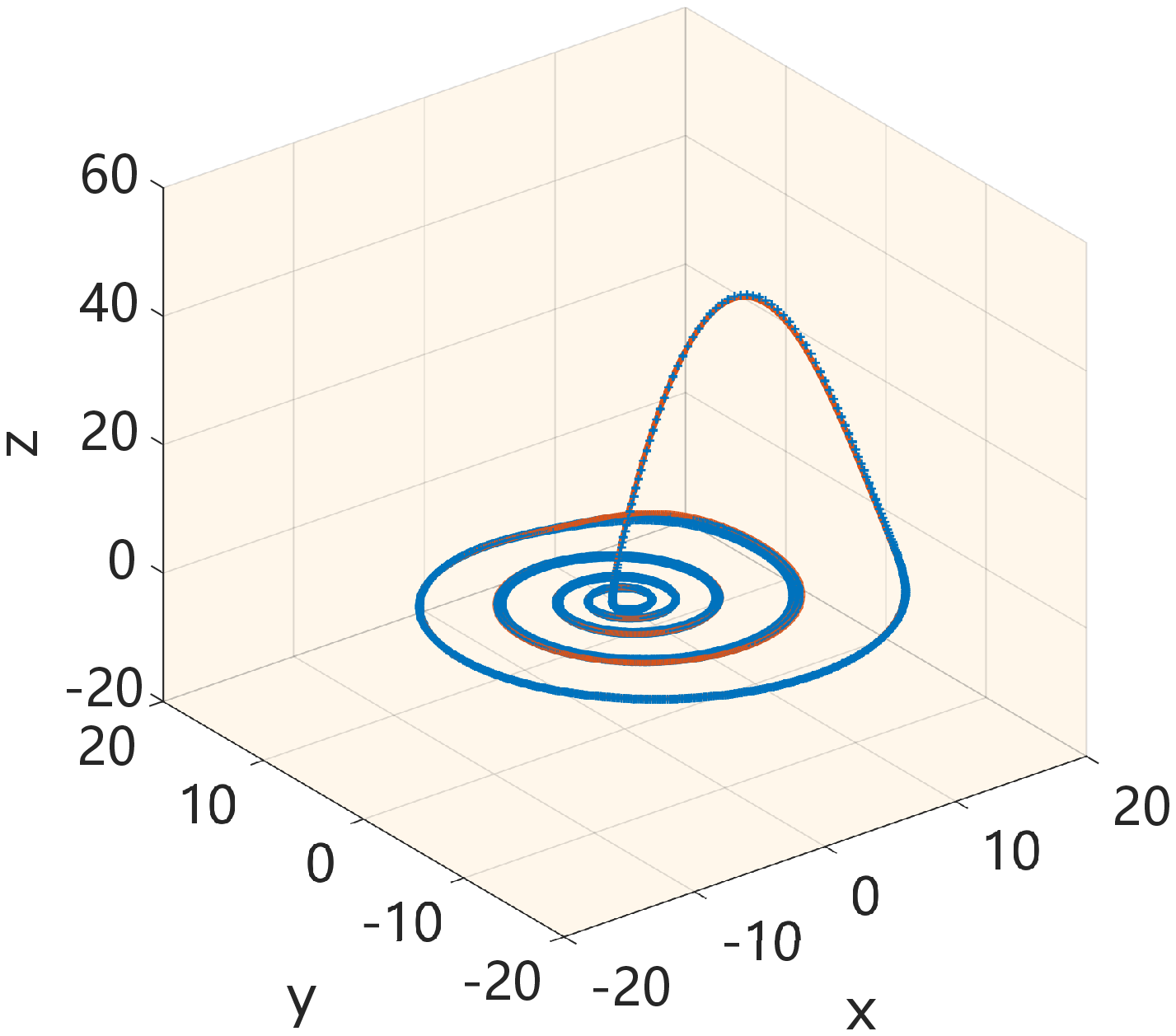}}
\subfigure[]{\includegraphics[width=4cm,height=4cm]{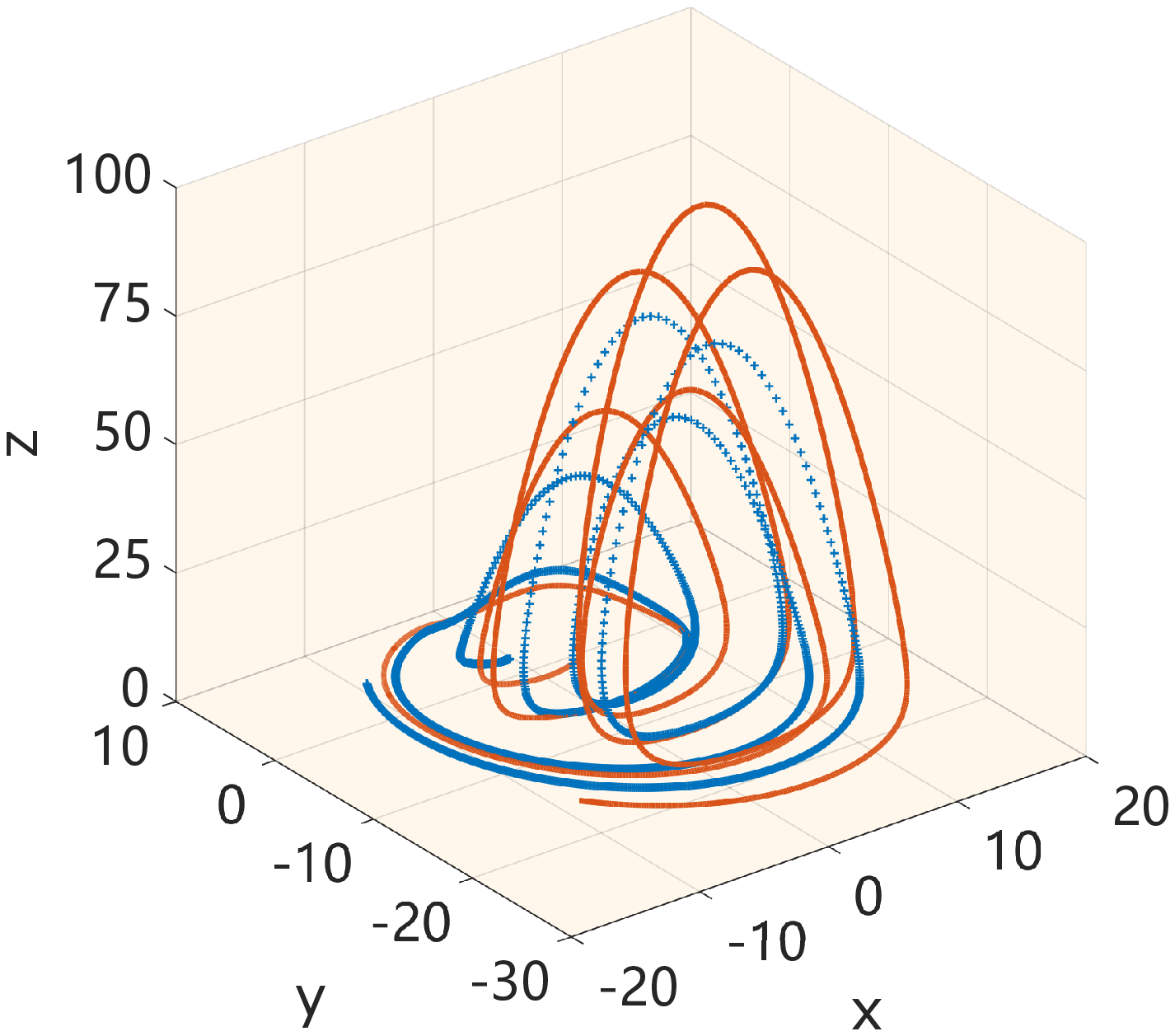}}
\subfigure[]{\includegraphics[width=4cm,height=4cm]{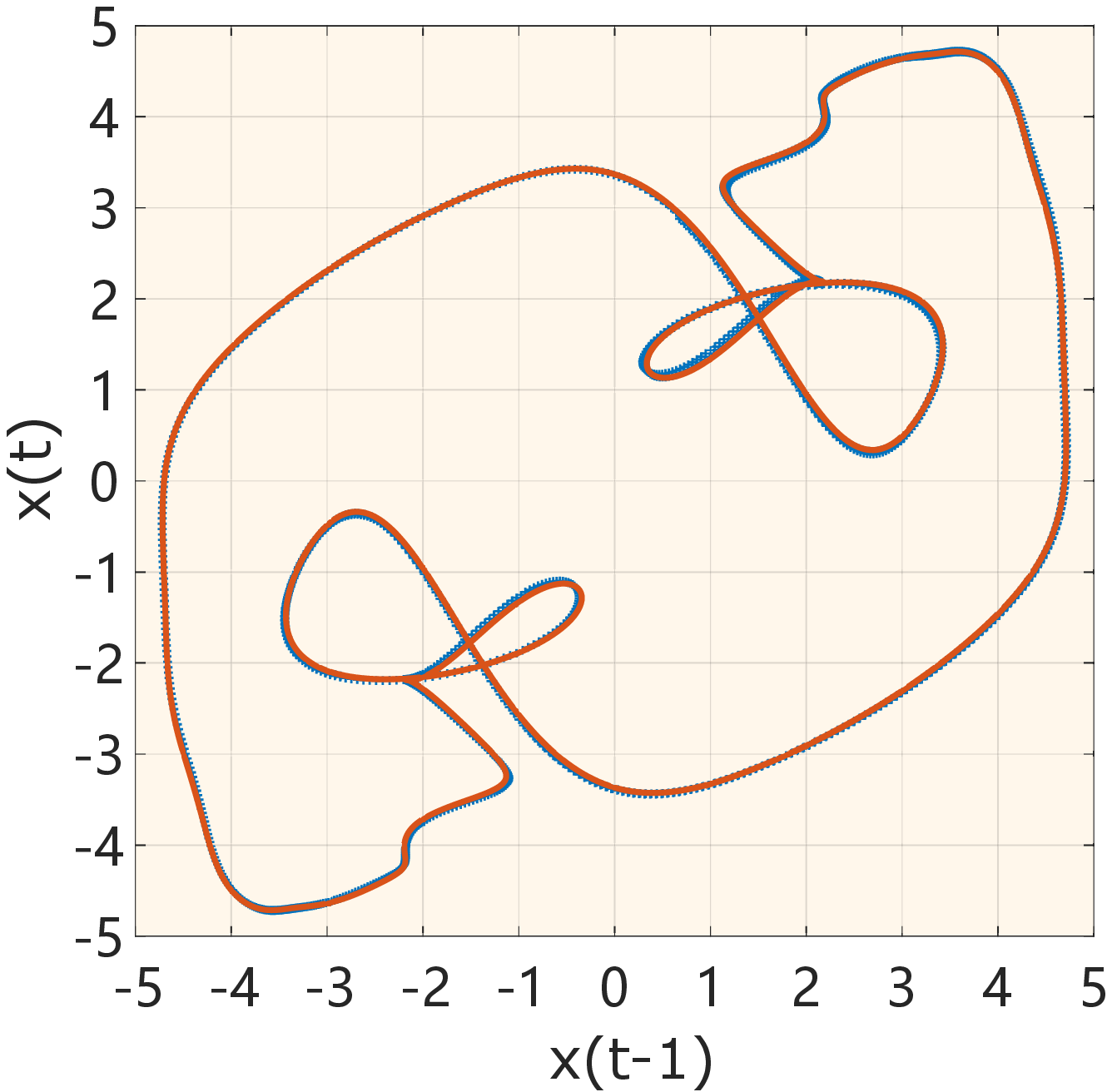}}
\caption{(a)-(d) are the exhibitions of the original system and the reconstructed system in experiments 1-4. In each subfigure, the blue dots stand for the
original system dynamic and the yellow solid line represents the reconstructed. (a)-(c) are shown in the space trajectories while (d) is in the phase portrait. It is shown that the successfully reconstructed systems reproduce the dynamics of the original systems. }
\end{figure}
\section{Discussion}
To understand the proposed method more precisely, some technical issues are discussed in this section.\par
\begin{itemize}
\item The total evaluation times is ${N_{out}\cdot I_{max}^{out}\cdot N_{in}\cdot I_{max}^{in}}$.
When the hyper-parameters are set according to Table II, 
the evaluation times becomes $M\cdot M^2\cdot Mn_p\cdot (Mn_p)^2$.
Afterwards, the complexity of the proposed method can be written as $O(n_p^3\cdot M^6)$.
$M$ is a given constant as illustrated in section II.
As a consequence, the complexity is rewritten as $O(n_p^3)$.
$n_p$ is influenced by the construction of the parameterized dictionary.
It barely increases as the complexity of the dictionary grows 
so that the complexity of the method could keep in a low level when faced with complex systems.
\end{itemize}
\begin{itemize}
\item The optimization capability of the method is related with the selection of EAs. 
Although only PSO is tested in the experiment, other EAs also have the equivalent ability in theory aspect.
Further, when specific algorithms are selected, the hyper-parameters need tuning in order to obtain 
better performance of optimization.
\end{itemize}
\begin{itemize}
\item Unsuccessful reconstruction is inevitable theoretically even when $N$ and $I_{max}$ in EAs are set big enough
due to the stochastic property of EAs. Hence, it is advised to perform the algorithm with proper times in practice.
\end{itemize}

\section{Conclusion}
In summary, this paper proposes a new data-driven method for reconstructing the system governing equation.
The details of the proposed method are illustrated.
Evolutionary computation is introduced to solve the reconstruction problem.
5 systems are tested including 3 chaotic DDEs.
The results show the effectiveness of the method.
Last but not least, this method is a generic method which is able to find other governing equations like
ODE and fractional differential equation(FDE).


%

\appendices




\ifCLASSOPTIONcaptionsoff
  \newpage
\fi



\bibliographystyle{IEEEtran}
\bibliography{IEEEabrv,IEEEtrans}
\end{document}